\documentclass{aa}

\usepackage{graphicx}
\usepackage{txfonts}

\usepackage{graphicx}	% Including figure files
\usepackage{amsmath}	% Advanced maths commands
\usepackage{xcolor}
\usepackage{hyperref}
\usepackage{placeins}

\newcommand{\units}[1]{\,\mathrm{#1}}
\newcommand{\msun}{{\rm M}_\odot}
\newcommand{\cbhbd}{$\tt{cBHBd}$\xspace}
\newcommand{\nthG}{\mathrm{n}^\mathrm{th}\mathrm{G}}

\begin{document}

\title{Repopulating the pair-instability mass gap without sustained growth to massive IMBHs: the case of 47\,Tuc}
\titlerunning{Repopulating the pair-instability mass gap: the case of 47~Tuc} % shows \textit{Gaia} BH3 formed as an exchange binary}

\author{Debatri Chattopadhyay\thanks{These authors contributed equally to this work.}\inst{,1,2,3}
\and
Daniel Marín Pina\footnotemark[1]\inst{,4}
\and
Mark Gieles\inst{5,6,7}
\and
Fabio Antonini\inst{3}
\and
Fotios Fronimos Pouliasis\inst{6}
}

\institute{
    Center for Interdisciplinary Exploration and Research in Astrophysics (CIERA) and Department of Physics \& Astronomy, Northwestern University, 1800 Sherman Ave, Evanston, IL 60201, USA\\
    \email{debatri.chattopadhyay@northwestern.edu}
    \and
    NSF-Simons AI Institute for the Sky (SkAI), 172 E. Chestnut St., Chicago, IL 60611, USA
    \and
    Gravity Exploration Institute, School of Physics and Astronomy, Cardiff University, Cardiff, CF24 3AA, United Kingdom
    \and
    Zentrum für Astronomie der Universität Heidelberg, Institut für Theoretische Astrophysik, Albert-Ueberle-Str. 2, 69120 Heidelberg  
    \email{daniel.marin@uni-heidelberg.de}
    \and
    ICREA, Pg. Llu\'is Companys 23, E08010 Barcelona, Spain
    \and
    Institut de Ciències del Cosmos (ICCUB), Universitat de Barcelona (IEEC-UB), Martí i Franquès 1, E08028 Barcelona, Spain
    \and
    Institut d'Estudis Espacials de Catalunya (IEEC), Edifici RDIT, Campus UPC, 08860 Castelldefels (Barcelona), Spain
}

\date{Received XX; accepted XX}
\abstract{We model the formation and retention of the most massive black hole (BH) in 47~Tuc using the semi-analytical code \texttt{cBHBd}, coupling cluster evolution with binary BH dynamics and computing merger-remnant masses, spins, and gravitational-wave recoil kicks via numerical-relativity surrogate prescriptions. We evolve 80\,000 cluster realisations spanning initial masses, densities, IMFs, and metallicities, in both a baseline scenario ($m_{\rm max} = 130\,\mathrm{M}_{\odot}$) and an extended-IMF scenario with ${\sim}\,50-110$ primordial BH seeds above the pair-instability gap ($M_{\rm BH} \sim 130-700\,\mathrm{M}_{\odot}$). Selecting models reproducing 47~Tuc's present-day mass and half-mass radius, we find hierarchical mergers alone yield a most massive retained BH of $M_{\rm BH} \sim 45-70\,\mathrm{M}_{\odot}$ with spin $\chi_{\rm BH} \sim 0.65$, limited to ${\sim}\,1-3$ mergers, as second-generation remnants acquire spin $\chi \sim 0.7$ that amplifies recoil kicks in subsequent generations. When primordial seeds are included, the retained-mass distribution becomes bimodal -- in ${\sim}\,90\%$ of realisations all seeds are ejected, but in ${\sim}\,10\%$ a massive seed ($M_{\rm BH} \gtrsim 450\,\mathrm{M}_{\odot}$) survives -- while the joint mass--spin distribution is trimodal; seeds surviving via stellar-mass BH mergers retain low spin ($\chi \lesssim 0.3$), whereas seed--seed mergers produce high-mass, high-spin remnants ($\chi \sim 0.65-0.7$), yielding 90th-percentile retained masses of ${\sim}\,500-1100\,\mathrm{M}_{\odot}$. Both scenarios are consistent with the $3\sigma$ dynamical upper limit of $578\,\mathrm{M}_{\odot}$. Our results favour a dark-remnant subsystem over a single massive IMBH and provide a spin--mass diagnostic testable with LIGO-Virgo-KAGRA, the Einstein Telescope, Cosmic Explorer, and LISA.}


\maketitle

%%%%%%%%%%%%%%%%%%%%%%%%%%%%%%%%%%%%%%%%%%%%%%%%%%

%%%%%%%%%%%%%%%%% BODY OF PAPER %%%%%%%%%%%%%%%%%%

% \dani{TODO: Not all authors have filled the acknowledgements}

\section{Introduction} \label{sec:Intro}
Intermediate-mass BHs (IMBHs; $10^{2-5}\units{\msun}$) occupy the mass range between stellar-mass BHs and the supermassive BHs in galactic nuclei \citep{Volonteri2010}. Establishing (or ruling out) IMBHs is important for understanding BH seeding and growth, the dynamical evolution of dense stellar systems, and the origin of compact-object merger populations observed through gravitational waves. Despite increasing observational hints for IMBH-like masses in some environments \cite[e.g. $\omega$~Cen,][]{Haberle2024}, robust and unambiguous detections remain challenging, particularly in star clusters where stellar remnants and mass segregation can mimic the dynamical signatures of a central massive object.

Gravitational wave observations by LIGO-Virgo-KAGRA (LVK) now map the merging BH mass spectrum and its high-mass tail, with population analyses indicating structure in the primary-mass distribution and a rapid decline at the highest masses, while events such as GW190521 produce remnants in the IMBH-mass regime \citep[e.g.][]{LIGOScientific:2020iuh,KAGRA:2021duu,LIGOScientific:2025slb}. From the stellar-evolution side, pair-instability processes are expected to suppress the formation of BHs in a characteristic `upper mass gap' scale; pair-instability theory predicts a dearth of BHs in a
characteristic mass range \citep{HegerWoosley2002,Woosley2017,
SperaMapelli2017}, and recent LVK population analyses find that
the primary-mass distribution extends into this region with features
consistent with a pre-existing gap repopulated by hierarchical
mergers accompanied by a high-spin component
\citep{Antonini2025arXiv250904637A}.

Dense star clusters provide two broad formation pathways to IMBHs. The first is a rapid channel, in which stellar collisions and runaway growth in very dense, young clusters can produce a massive BH seed \citep{Portegies2002, Portegies2004, Vergara2023, Vergara2025}. The second is a hierarchical-merger channel in which repeated mergers of stellar-mass BHs build up a more massive remnant over time \citep{Oleary2006ApJ, Antonini2016, Antonini:2018auk}. Earlier work by \citet{Shapiro1985ApJ} 
and \citet{Quinlan1987ApJ} explored BH seed
formation through relativistic collapse in very dense clusters.

A key obstacle to hierarchical growth is gravitational wave recoil: asymmetric emission of gravitational radiation imparts a kick to the merger remnant that can exceed the cluster escape speed, removing the growing BH from the system \citep{HolleyBockelmann2008ApJ, Gerosa2019PhRvD}. 
The kick amplitude depends on both the progenitor mass ratio
and the spins \citep{BakerKick2007, GonzKick2007, CampanelliKick2007,LoustoKick2011}. This naturally connects the feasibility of IMBH growth to the cluster’s escape velocity and to the evolving mass and spin distribution of the BH population \citep{Mapelli2016, Gerosa2021NatAs,DiCarlo2021, Chattopadhyay2023MNRAS}.

47~Tucanae (47~Tuc, NGC~104) is among the most massive globular clusters in the Milky Way and has long been considered a promising host for an IMBH. However, current dynamical constraints yield only upper limits on any central BH mass, with recent modelling placing a stringent $3\sigma$ limit of $M_{\rm BH}<578\units{\msun}$ \citep{DellaCroce2024}. Moreover, the cluster’s kinematics can be reproduced without an IMBH if a population of stellar-mass BHs (about dozens, with total mass $\sim$$10^{2-3}\units{\msun}$) is retained in the core  \citep{HenaultBrunet2020}, motivating renewed interest in scenarios where 47~Tuc is dominated by a dark-remnant subsystem rather than a single central object. Previous Monte Carlo simulations of 47\,Tuc by \citet{GierszHeggie2011} using the \texttt{MOCCA} code \citep{Hypki2013} predicted the retention of $\sim 17$ stellar-mass BHs at 12\,Gyr with no IMBH, broadly consistent with the dark-remnant subsystem
interpretation.

Recent ultra-deep radio imaging with ATCA has revealed a faint compact source coincident with the photometric centre of 47\,Tuc \citep{Paduano2024},
consistent with -- but not uniquely attributable to -- an accreting BH of mass
$\sim 54 - 6\,000\units{\msun}$. Independently, theoretical models of cluster
formation predict that proto-clusters as massive as 47\,Tuc hosted extremely
massive stars ($\gtrsim 10^3\units{\msun}$) whose direct collapse could have
seeded a population of `lite' IMBHs ($\sim 130 - 700\units{\msun}$) above
the pair-instability gap \citep{Gieles2025_inertialInflows, Padoan2020}.
Whether such seeds survive to the present epoch depends on the interplay
between gravitational-wave recoil kicks, dynamical ejection, and the cluster's escape speed -- a
question that the semi-analytical framework of \texttt{cBHBd} is well suited
to address.

In this work, we quantify the mass and spin ranges of the most massive BH that can be produced and retained under plausible 47~Tuc-like initial conditions. We evolve a large suite of semi-analytical cluster models using \texttt{cBHBd}, varying the initial cluster mass and density, the initial mass function (IMF), and metallicity over ranges motivated by the uncertainties in 47~Tuc’s early formation environment. We explicitly follow hierarchical BH mergers and compute remnant properties with numerical-relativity (NR) surrogate prescriptions, enabling a more realistic treatment of spin-dependent recoil and its impact on retention. We then identify the subset of models that reproduce 47~Tuc’s present-day global properties and use these to infer the most likely mass and spin of the cluster’s most massive retained BH.

The paper is organised as follows: Section~\ref{sec:methods} describes the cluster models and the remnant prescriptions. Section~\ref{sec:results} presents the distributions of the massive BH masses and spins retained and ejected, and discusses their dependence on initial cluster properties. 
Section~\ref{sec:discussion} discusses the implications of the results for interpreting 47~Tuc's dynamical constraints, and Section~\ref{sec:conclusions} summarises our main conclusions.
% Sections~\ref{sec:discussion} and \ref{sec:caveats} summarise our conclusions and implications for interpreting 47~Tuc’s dynamical constraints.
% \dani{move caveats to appendix?}

\section{Methods}
\label{sec:methods}

We use the semi-analytical fast code \cbhbd \citep{Antonini:2019ulv, Antonini:2022vib}, with the updates of Fronimos Pouliasis et al. (2025, submitted)%\citep{FronimosPouliasis2025} \dani{Fotis' paper, to be published}
, to simulate the evolution of 47~Tuc-like clusters. \cbhbd is based on H\'enon's principle \citep{Henon:1972} of balanced evolution, which states that, after an initial evolution phase, the energy required for the cluster evolution is powered by the external energy created at the cluster core. By assuming that the energy creation is due to the formation and tightening of binary BHs (BBH), \cbhbd couples the cluster's properties with its internal BBH dynamics, and allows one to recover BBH mergers and their associated IMBH formation.

The initial mass function is evolved through the single stellar evolution code $\tt{SSE}$ by \cite{HurleySSE:2000pk}, with updated wind mass loss prescription \citep{Vink2001}, (pulsational) pair instability or (P)PISN recipe \citep{SperaMapelli2017} and supernova kick Maxwellian distribution (with $\sigma=265\units{km/s}$) from \cite{Hobbs:2005}; scaled by fallback mass for BHs \citep{Fryer:2012, Belczynski:2010}.

For this work, we have updated \cbhbd to compute the properties of the merger remnant of a BBH (mass, spin, and recoil kick) using state-of-the-art surrogate models calibrated with non-equal-mass NR simulations. We use the \texttt{NRSur7dq4Remnant} model \citep{Varma2019} for mergers with a mass ratio $q\leq 6$ (where $q=m_1/m_2$ and $m_2\leq m_1$). This model accounts for the effects of masses and individual spin magnitudes and orientations, but accuracy is not guaranteed at higher mass ratios. Instead, for mergers with $q>6$, we use the \texttt{BHPTNRSurRemnant} model \citep{Islam2023}, which can be extrapolated up to $q\sim 1000$. All first-generation BHs are assigned a natal dimensionless spin
$\chi_{\rm natal} = 0$, consistent with efficient angular-momentum
transport in massive stellar progenitors \citep{FullerMa2019}. Spin
is subsequently acquired only through binary mergers.

% \dani{i can write more on the methods if we have the space}\debatri{all good for now unless the referee wants}

\subsection{Base Models} \label{sec:base_models}
We evolve a total of 40\,000 models with 40 unique initial cluster mass, density, initial mass function and metallicity set-ups, each re-modelled a 1000 times to account for statistical fluctuations. 
We have two values for the initial cluster mass, $M_\mathrm{cl,i}=[2, 4] \times 10^6\units{\msun}$, five for the cluster initial half-mass density, $\rho_\mathrm{h,i}=[10^3,3\times 10^3,10^4,3\times 10^4, 10^5]\units{\msun}\units{pc^{-3}}$ \citep[guided broadly by Monte-Carlo simulations by][]{Ye2022}, and two different initial mass functions: \cite{Kroupa2001}, denoted by `\texttt{K}' in the model names, and \cite{HenaultBrunet2020} -- more realistic for 47~Tuc -- denoted by `\texttt{H}' in the model names. We also inspect the effect of two different metallicities, $Z=[0.003,0.007]$ (denoted by `\texttt{Z1}' and `\texttt{Z2}' in the model names as suffixes), accounting for the range of metallicities observationally reported for 47~Tuc \citep{Pasquini1997,Koch2008}, including the evidence of metallicity spread in the cluster \citep[][]{PopSpreadCriscienzo2010,MultPopVentura2014,MultiplePopulationLee2022}. We expect `\texttt{Z1}' and `\texttt{Z2}' variations to show extremum limits to the mass of BH formed through hierarchical mergers. All cluster models are evolved for an age sampled uniformly between 10.4\,Gyr to 13.4\,Gyr \citep{AgeGibson1999,Brogaard2017}, in a circular orbit around a Milky Way-like galaxy with an effective Galactocentric radius of $R_\mathrm{G}=7.4\units{kpc}$, approximating the mildly eccentric ($\approx0.16$) orbit of 47\,Tuc \citep{harris2010new}. Our models' initial conditions are summarised in Table~\ref{tab:tableModel}. % \ff{Should we mention also an effective $R_G$ used for simulating those in circular orbits, or where they simulated as isolated?}

\subsection{Seed Models} \label{sec:seed_models}

% \dani{Use Spera Mapelli for the map of progenitor masses to BH masses, bare BSE does not work}
Our baseline models assume a maximum stellar mass of $m_{\rm max} = 130\units{\msun}$, which, after stellar evolution with included prescriptions of (P)PISN mass-gap,
yields first-generation BHs with masses $\lesssim 45\units{\msun}$ (at $Z=0.007$) to
$\lesssim 55\units{\msun}$ (at $Z=0.003$). However, recent theoretical work on
globular cluster formation predicts that proto-clusters hosted extremely massive
stars (EMSs) with masses $\gtrsim 10^3\units{\msun}$
\citep{Gieles2025_inertialInflows}, motivated both by the inertial-inflow
theory of massive star formation \citep{Padoan2020} and by the need to
explain the ubiquitous light-element abundance anomalies in globular clusters
(see \citealt{Bastian2018} for a review). We note that while stellar-mass BHs that undergo subsequent mergers are sometimes loosely referred to as seeds in the literature, we reserve the term `seeds' here exclusively for primordial BHs above the (P)PISN gap.

Specifically, \citet{Gieles2025_inertialInflows} predict that the stellar
mass function in a 47\,Tuc-like proto-cluster ($M_{\rm cl,i} \sim 2 \times
10^6\units{\msun}$) at the end of the formation epoch ($t \approx 2.2$\,Myr)
contains $\sim 6$ stars with masses $> 10^3\units{\msun}$ and $\sim 130$ stars
with masses $> 250\units{\msun}$. These numbers reflect the raw stellar mass
function before wind mass loss reduces the stellar masses prior to
core collapse; the actual number of BH seeds above the (P)PISN
gap depends on the adopted wind prescription and the evolution of the star and its core mass.

We convert the predicted stellar mass function into an initial BH seed
population as follows. For each progenitor star with zero-age main-sequence
(ZAMS) mass $M_\star$ at $t=2.2$\,Myr, we apply metallicity-dependent wind
mass loss using the \citet{Vink2001} prescription over the remaining
main-sequence lifetime ($\sim 1 - 2$\,Myr for $M_\star \gtrsim 250\,
\units{\msun}$), obtaining a pre-collapse mass $M_{\rm pre}$. The helium core mass
at collapse is approximately $M_{\rm He} \approx (0.50 - 0.60) \times
M_{\rm pre}$, depending on metallicity and mixing assumptions
\citep{SperaMapelli2017}. Stars whose helium cores fall in the range
$64\units{\msun} \lesssim M_{\rm He} \lesssim 133\units{\msun}$ are completely
destroyed by pair-instability supernovae and leave no remnant
\citep{HegerWoosley2002, Woosley2017}. Stars with $M_{\rm He} >
133\units{\msun}$ bypass the (P)PISN regime via photodisintegration and collapse
directly into BHs, retaining nearly all of their pre-collapse mass minus
$\sim 10\%$ neutrino losses \citep{HegerWoosley2002}. We adopt a simplified
IFMR for the direct-collapse (DC) regime:
\begin{equation}\label{eq:ifmr}
    M_{\rm BH} \approx f_{\rm DC}(Z) \times M_{\star,\,2.2\,\rm Myr} \;,
\end{equation}
where $f_{\rm DC}$ is the net retention fraction accounting for wind mass
loss and neutrino losses. We adopt $f_{\rm DC} = 0.6$ for $Z = 0.003$ and
$f_{\rm DC} = 0.45$ for $Z = 0.007$, reflecting the stronger wind mass loss
at higher metallicity \citep{Vink2001, SperaMapelli2017}.

The (P)PISN gap filter eliminates progenitors whose final helium cores fall inside
the $64 - 133\units{\msun}$ window. At $Z = 0.003$, this removes progenitors
with $M_{\star,\,2.2\,\rm Myr} \approx 250 - 320\units{\msun}$ (whose reduced
pre-collapse masses yield $M_{\rm He}$ within the gap). At $Z = 0.007$,
the stronger winds shift this exclusion window upward to $M_{\star,\,2.2\,
\rm Myr} \approx 250 - 400\units{\msun}$. Stars above these thresholds produce
BHs via direct collapse; stars below produce BHs through the standard
(P)PISN-limited channel already handled by \texttt{SSE}.

Applying this procedure to the \citet{Gieles2025_inertialInflows} mass
function yields the following seed populations:

\begin{enumerate} 

    \item \textbf{$Z = 0.003$ (Z1 models):} We inject $N_{\rm seed} \approx 
    80 - 110$ BH seeds above the (P)PISN gap, with masses drawn from a truncated
    power law $\mathrm{d}N / \mathrm{d}M_{\rm BH} \propto M_{\rm BH}^{-2.3}$
    between $M_{\rm BH,min} = 130\units{\msun}$ and $M_{\rm BH,max} =
    700\units{\msun}$. The most massive $\sim 3 - 5$ seeds have $M_{\rm BH}
    \gtrsim 500\units{\msun}$, corresponding to the $\sim 6$ progenitors above
    $10^3\units{\msun}$.
    \item \textbf{$Z = 0.007$ (Z2 models):} We inject $N_{\rm seed} \approx 
    50 - 75$ seeds, with the same power-law slope but $M_{\rm BH,min} =
    130\units{\msun}$ and $M_{\rm BH,max} = 500\units{\msun}$. The most massive
    $\sim 2 - 4$ seeds have $M_{\rm BH} \gtrsim 350\units{\msun}$.
\end{enumerate}

These seed populations are injected in addition the standard BH mass
spectrum generated by SSE (which handles progenitors up to $m_{\rm max} =
130\units{\msun}$ as before). The seeds are added directly to the initial BH mass
bins in \texttt{cBHBd} at $t = 0$, bypassing \texttt{SSE}; this is justified because
EMSs complete their evolution within $\lesssim 3$\,Myr, well before the
onset of balanced cluster evolution. All seeds are assigned an initial
dimensionless spin of $\chi_{\rm natal} = 0$, appropriate for BHs formed via
direct stellar collapse rather than hierarchical mergers \citep{FullerMa2019}.

Each realisation draws its seed masses independently from the power-law
distribution described above, so that stochastic variation in the number and
masses of the most massive seeds propagates naturally through the statistics.
We retain all other model parameters – $M_{\rm cl,i}$, $\rho_{h,i}$, IMF (Kroupa or H\'enault-Brunet), 
metallicity, age, and Galactocentric orbit – identical to the baseline grid, and evolve 1000 
realisations per model, yielding 40\,000 new runs. Combined with the 40\,000 baseline realisations 
of Sec.\ref{sec:base_models}, the full simulation suite comprises 80\,000 runs in total. The full model
grid is repeated (rather than restricting to analogues) so that the analogue
selection criteria of Sec.~\ref{sec:clusterMassRadius} can be applied
a~posteriori on the same footing as the baseline models. We denote
these extended-IMF models with the suffix `S' (for `seeded'), e.g.\
\texttt{M08KZ1S}; their initial conditions are summarised in Table~\ref{tab:tableModel_S}.

\section{Results}
\label{sec:results}

\subsection{Cluster mass and radius} 
\label{sec:clusterMassRadius}
Observationally constrained current dynamical models of 47~Tuc consistently place its present-day mass in the range
$M_{\rm cl,f} \simeq (0.78-1.06) \times 10^{6}\units{\msun}$ and its three-dimensional half-mass radius in the range
$r_{\rm h,f} \simeq 5.6-8.2\,{\rm pc}$.
For example, \citet{Baumgardt2018MNRAS} infer $M_{\rm cl,f}=7.79\times 10^{5}\units{\msun}$ and $r_{\rm h,f}=5.62\,{\rm pc}$ from $N$-body-based fits to surface-density and kinematic profiles, while \citet{HenaultBrunet2020} find $M_{\rm cl,f}=1.06\times 10^{6}\units{\msun}$ and $r_{\rm h,f}=8.16\,{\rm pc}$ from multi-mass dynamical modelling. A recent multi-mass model fit by \citet{Dickson2023MNRAS} gives $M_{\rm cl,f}=8.94\times 10^{5}\units{\msun}$ and $r_{\rm h,f}=6.69\,{\rm pc}$.

\begin{figure}
    \resizebox{\hsize}{!}{\includegraphics{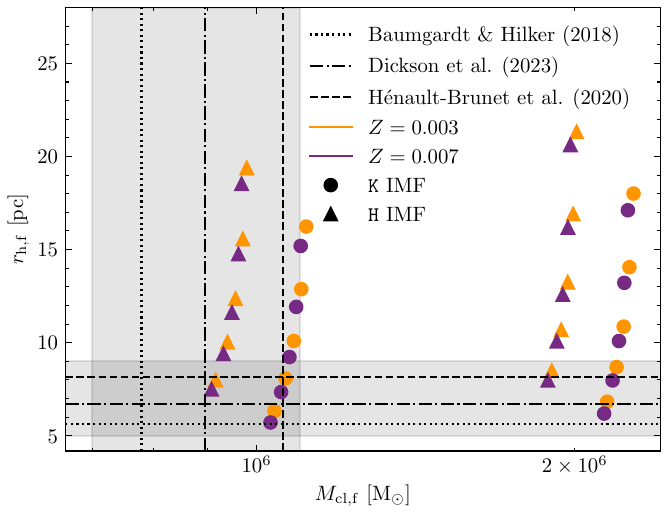}}
    \caption{Scatter plot of the median half-mass radius ($r_\mathrm{h, f}$) versus the median cluster mass ($M_\mathrm{cl, f}$), both measured at the present time, for each model in our simulations. The points represent the median over 1000 runs. The models are separated by metallicities (marker colour) and IMF (marker shape). The dotted, dashed, and dash-dotted lines black lines are the values of $M_\mathrm{cl, f}$ and $r_\mathrm{h,f}$ found in the cited papers. In shaded grey, the ranges $5\leq r_\mathrm{h}/\mathrm{pc}\leq 9$ and $7\times 10^5\leq M_\mathrm{cl, f}/\msun \leq1.1\times 10^6$ %$\pm0.03\units{dex}$ over the \cite{HenaultBrunet2020} value for $M_\mathrm{cl, f}$
    , which serve as a heuristic cutoff to delimit which of our models match the current properties of 47~Tuc (i.e. `47~Tuc analogues'). These six base 47~Tuc analogues are \texttt{M04KZ1}, \texttt{M05KZ1}, \texttt{M04KZ2}, \texttt{M05KZ2}, \texttt{M05HZ1} and \texttt{M05HZ2}; their six seeded counterparts also satisfy the cut-offs, making a total of 12 unique models (each re-modelled a 1000$\times$).}
    \label{fig:Mclf-rhf}
\end{figure}

We illustrate the present-day cluster mass and half-mass radius in Fig.~\ref{fig:Mclf-rhf}. Under our assumed external tidal field, some models with $M_{\rm cl,i}=2\times10^{6}\units{\msun}$ (base models \texttt{M01}, \texttt{M02}, \texttt{M03}, \texttt{M04}, \texttt{M05}; across both IMF and metallicity choices) evolve to present-day masses consistent with the observationally inferred $M_{\rm cl,f}$ of 47~Tuc within our adopted tolerance band of $7\times 10^5\leq M_\mathrm{cl, f}/\msun \leq1.1\times 10^6$%$\pm 0.03\units{dex}$ around the value of \cite{HenaultBrunet2020}
. In contrast, models with $M_{\rm cl,i}=4\times10^{6}\units{\msun}$ (base models \texttt{M06}, \texttt{M07}, \texttt{M08}, \texttt{M09}, \texttt{M10}) remain systematically more massive at late times, indicating insufficient mass loss in our baseline tidal treatment to reconcile them with the present-day cluster mass.

A similar selection can be performed using the present-day half-mass radius. Motivated by the range of values quoted above, we adopt $5 \le r_{\rm h,f}/{\rm pc} \le 9$ as a conservative acceptance window. In Fig.~\ref{fig:Mclf-rhf}, this immediately excludes a set of models whose median half-mass radii are clearly too extended, with $r_{\rm h,f}>10$~pc. In particular, the models \texttt{M01KZ1}, \texttt{M02KZ1}, \texttt{M03KZ1}, \texttt{M06KZ1}, \texttt{M07KZ1}, and \texttt{M08KZ1}, and \texttt{M01HZ1}, \texttt{M02HZ1}, \texttt{M03HZ1}, \texttt{M04HZ1}, \texttt{M06HZ1}, \texttt{M07HZ1}, \texttt{M08HZ1}, and \texttt{M09HZ1} (for $Z=0.003$) and \texttt{M01KZ2}, \texttt{M02KZ2}, \texttt{M06KZ2}, \texttt{M07KZ2}, and \texttt{M08KZ2}, and \texttt{M01HZ2}, \texttt{M02HZ2}, \texttt{M03HZ2}, \texttt{M06HZ2}, \texttt{M07HZ2}, \texttt{M08HZ2}, and \texttt{M09HZ2} (for $Z=0.007$) have medians at or beyond 10~pc, with several lying well above $\sim12-18\units{pc}$. We therefore treat these models as definitively inconsistent with the present-day size of 47~Tuc in our baseline setup.

% \texttt{M01KZ1}, \texttt{M02KZ1}, \texttt{M03KZ1}, \texttt{M06KZ1}, \texttt{M07KZ1}, \texttt{M08KZ1}, \texttt{M01HZ1}, and \texttt{M06HZ1} (for $Z=0.003$) and \texttt{M01KZ2}, \texttt{M02KZ2}, \texttt{M08KZ2}, \texttt{M06KZ2}, \texttt{M07KZ2}, \texttt{M01HZ2}, and \texttt{M06HZ2} (for $Z=0.007$) have medians at or beyond 10~pc, with several lying well above $\sim12-18\units{pc}$. We therefore treat these models as definitively inconsistent with the present-day size of 47~Tuc in our baseline setup.

Combining this half-mass radius criterion with the present-day mass constraint yields a small subset of models that simultaneously reproduce both the mass and size of 47~Tuc in our baseline setup. Specifically, the base models \texttt{M04KZ1}, \texttt{M05KZ1}, \texttt{M04KZ2}, and \texttt{M05KZ2} fall within the accepted windows for both $M_{\rm cl,f}$ and $r_{\rm h,f}$, and we therefore adopt these as our primary `47~Tuc-analogue' realizations. Their seeded counterparts (\texttt{M04KZ1S}, \texttt{M05KZ1S}, \texttt{M04KZ2S}, \texttt{M05KZ2S}) inherit the same initial cluster properties and are adopted as the seeded analogues. Among the corresponding H\'enault-Brunet IMF models at the same densities, \texttt{M05HZ1} and \texttt{M05HZ2} (as well as the seeded counterparts \texttt{M05HZ1S} and \texttt{M05HZ2S}) also satisfy both criteria, whereas \texttt{M04HZ1} and \texttt{M04HZ2} remain a little too extended in half-mass radius. This does not imply that either IMF is physically preferred for 47~Tuc; rather, the analogue selection reflects a degeneracy between the IMF and initial structural parameters under our simplified tidal treatment.

% The corresponding H\'enault-Brunet IMF models at the same densities
% (\texttt{M04HZ1}, \texttt{M05HZ1}, \texttt{M04HZ2},
% \texttt{M05HZ2}) do not simultaneously satisfy both criteria: the
% top-heavier IMF drives stronger early mass loss, shifting these
% models outside the accepted windows. This does not imply the Kroupa
% IMF is physically preferred for 47~Tuc; rather, the analogue
% selection reflects a degeneracy between the IMF and initial
% structural parameters under our simplified tidal treatment.

We emphasise that these cuts are intended as a pragmatic filter rather than a formal exclusion. First, the mapping between an observed projected half-light radius and a three-dimensional half-mass radius is not unique: mass segregation and metallicity-dependent stellar evolution can bias light-based size estimates relative to mass-based sizes at the $\sim 10 - 20\%$ level (e.g. \citealt{Jordan2004}; see also the discussion in \cite{Mackey2005} and direct $N$-body experiments in \cite{Sippel2012MNRAS} and \cite{Schulman2012}), although we note that the multi-mass dynamical fits of \cite{HenaultBrunet2020} and \cite{Dickson2023MNRAS} already account for mass segregation in their reported half-mass radii. Second, the late-time mass-loss rate and equilibrium size depend on the details of the external tidal field. Time-dependent tides, orbital eccentricity, and disk/bulge shocks can enhance stripping and alter the structural evolution compared to a static, circular-orbit approximation \citep{Gnedin1997ApJ,BaumgardtMakino2003MNRAS,Miholics2014}. Consequently, models that marginally fail to meet our present-day windows could be brought into agreement under a stronger and/or evolving Galactic potential; we interpret the present selection accordingly.

\subsection{IMBH mass in non-seeded models}
\label{subsec:imbh_mass}

\begin{figure}
    \resizebox{\hsize}{!}{\includegraphics{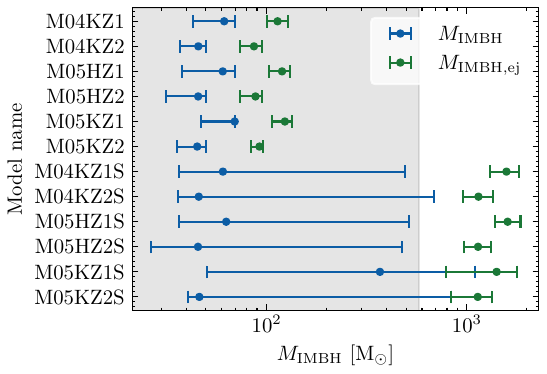}}
    \caption{Mass of the most massive BH retained ($M_\mathrm{IMBH}$) and ejected ($M_\mathrm{IMBH, ej}$) from the cluster, for each of our 47~Tuc-analogue models. The points and the widths of the errorbars represent the median, 10th and 90th percentiles over $1000$ runs. In gray, the limit imposed by \cite{DellaCroce2024} of $M_\mathrm{IMBH}< 578\units{\msun}$.}
    \label{fig:mIMBH}
\end{figure}

\begin{figure}
  \resizebox{\hsize}{!}{\includegraphics{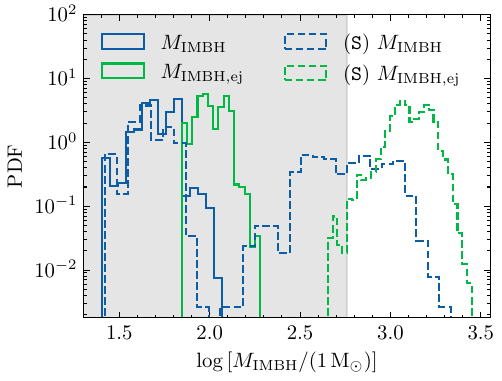}}
  \caption{Probability distribution function of the mass of the most massive BH remaining in the cluster ($M_{\rm IMBH}$, in blue) and the most massive ejected BH ($M_{\rm IMBH,ej}$, in green) for 47\,Tuc analogue models. Solid lines show the baseline (non-seeded) models; dashed lines show the seeded (\texttt{S}) models. For the baseline analogues, all retained masses lie well below the $3\sigma$ upper limit of $M_{\rm IMBH} < 578\units{\msun}$ from \citet{DellaCroce2024}; for the seeded analogues, the 90th-percentile values approach or exceed this limit, reaching ${\sim}\,500 - 1100\,\mathrm{M}_\odot$.}
  \label{fig:mIMBHspectra}
\end{figure}

\begin{figure}
    \resizebox{\hsize}{!}{\includegraphics{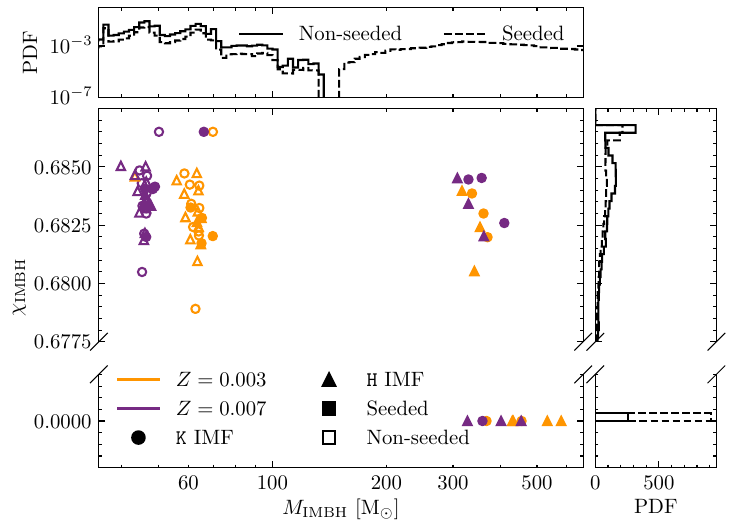}}
    \caption{Scatter plot of the mass ($M_\mathrm{IMBH}$) versus spin ($\chi_\mathrm{IMBH}$) of the most massive BH retained in the cluster at the end of the simulation, for each model in our simulation. The points represent the median over 1000 runs. The models are separated by metallicities (marker colour) and IMF (marker shape). On the top and right, the probability distribution functions of $M_\mathrm{IMBH}$ and $\chi_\mathrm{IMBH}$ across all runs.}
    \label{fig:mIMBH-chiIMBH}
\end{figure}

\begin{figure}
    \resizebox{\hsize}{!}{\includegraphics{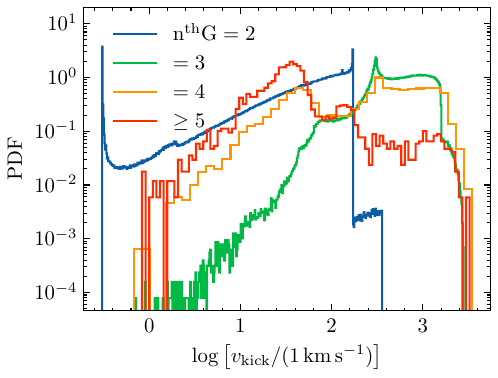}}
    \caption{Probability distribution function of the gravitational-wave recoil kick velocity after a BBH merger ($v_\mathrm{kick}$) in our 47\,Tuc analogue simulations, separated by the merger generation of the remnant BH ($\nthG$).}
    \label{fig:vkick}
\end{figure}
Having identified the 47~Tuc-analogue models, we now examine the masses and spins of the most massive BHs they produce. For each realisation, we record (i) the mass of the most massive BH remaining bound to the cluster at the present time, $M_{\rm IMBH}$, and (ii) the mass of the most massive BH ejected from the cluster by the present time, $M_{\rm IMBH,ej}$. We retain this notation for continuity with the literature; however, in our model grid $M_{\rm IMBH}$ and $M_{\rm IMBH, ej}$ should be understood as shorthands for the mass of the most massive BH produced by the cluster (retained or ejected, respectively), which in many realizations remains in the high stellar-mass or low-IMBH regime and should not be interpreted as a definitive `classical' IMBH. For each cluster model we summarize the distributions of these quantities using the median and the 10th--90th percentile range across realizations (Tables~\ref{tab:tableModel},~\ref{tab:tableModel_S}; Figs. \ref{fig:mIMBH}, \ref{fig:mIMBHspectra}, and \ref{fig:mIMBH-chiIMBH}).

Figures~\ref{fig:mIMBH} and \ref{fig:mIMBHspectra} show that the retained most-massive BH masses are modest across models without seeds: medians lie in the range $\sim 45-70\units{\msun}$, with 90th-percentile values $\lesssim 70\units{\msun}$.  All models therefore predict $M_{\rm IMBH}\ll 578\units{\msun}$, well below the current dynamical upper limit reported by \cite{DellaCroce2024}. The weak dependence of $M_{\rm IMBH}$ on initial cluster properties is reflected in Table~\ref{tab:tableModel}: increasing the initial escape velocity at the half-mass radius, $v_{\rm esc,h,i}$ (and, typically, the initial density), raises the typical merger generation of the most massive retained BH,
$\nthG$, only marginally, from $\nthG \simeq 1.7$ to $2.5$,
although these differences are comparable to the statistical scatter
and the distributions overlap substantially. 

This indicates that growth is usually
limited to one or two merger steps before the remnant is lost to
recoil, largely independent of the initial cluster properties within
the range explored, consistent with the low-$v_{\rm esc}$ plateau
found in \citet{Antonini:2018auk} and
\citet{Chattopadhyay2023MNRAS}. As shown in Fig.\ref{fig:vkick}, the recoil kick distribution shifts to progressively higher velocities and broader tails with each merger generation, making retention increasingly unlikely beyond the second or third generation.%\debatri{Dani: it's 47Tuc analogue models right?}\dani{no, it's all models}

The most massive ejected BHs are systematically heavier than the most massive retained BHs.  As shown in Figs.~\ref{fig:mIMBH} and \ref{fig:mIMBHspectra}, $M_{\rm IMBH,ej}$ has medians spanning $\sim 90-120\units{\msun}$, with the upper tail extending to $\sim 130\units{\msun}$ in the most favourable models.  This behaviour is expected if the cluster occasionally forms a relatively massive hierarchical merger remnant but fails to retain it, as the large gravitational wave recoil exceeds the cluster's escape speed.  In practice, for the escape-speed range explored here ($v_{\rm esc,h,i}\lesssim 171\units{km}\units{s^{-1}}$; Table~\ref{tab:tableModel}), retention of repeated merger remnants is inefficient, so the cluster is more likely to eject the largest remnants rather than retaining them.

Metallicity is the clearest systematic driver of the mass scale.  The $Z=0.007$ models yield smaller $M_{\rm IMBH}$ and $M_{\rm IMBH,ej}$ than the $Z=0.003$ models (Table~\ref{tab:tableModel}), consistent with stronger winds at higher metallicity producing lighter BH progenitors and hence lighter merger remnants.  For example, typical retained masses decrease from $\sim 60-70\units{\msun}$ at $Z=0.003$ to $\sim 45-50\units{\msun}$ at $Z=0.007$, while typical ejected masses decrease from $\sim 100-130\units{\msun}$ to $\sim 70-100\units{\msun}$.  The choice of IMF introduces a secondary effect, with the \cite{HenaultBrunet2020} IMF producing slightly larger remnant masses and slightly higher $\nthG$ at fixed $M_{\rm cl,i}$, $\rho_{\rm h,i}$, and $Z$, but the overall mass scale remains far below the observational IMBH limit. Thus, even among models tuned to match 47~Tuc’s present-day global properties, the most likely outcome is that the cluster does not retain an IMBH in the classical sense; instead, it retains a comparatively massive stellar-origin BH and ejects the most massive hierarchical-merger products.

% Finally, we focus on the subset of 47~Tuc-analogue models that satisfy both the present-day mass and size cuts defined in Sec.~\ref{sec:clusterMassRadius}. 
% Fig.~\ref{fig:mIMBHspectra} shows their combined distributions of $M_{\rm IMBH}$ and $M_{\rm IMBH,ej}$. The retained-mass distribution is broad but peaks at $M_{\rm IMBH}\sim 40-60\units{\msun}$, with a median $M_{\rm IMBH}=50\units{\msun}$, while the most-massive ejected remnants are shifted to larger masses and peak at $\sim 70-110\units{\msun}$ with a median $M_{\rm IMBH,ej}=100\units{\msun}$. For this subset, the lower metallicity $Z1$ models have $M_{\rm IMBH}\sim 60-70\units{\msun}$ and $M_{\rm IMBH, ej}\sim 100-120\units{\msun}$, while higher metallicity $Z2$ models have $M_{\rm IMBH}\sim 45\units{\msun}$ and $M_{\rm IMBH, ej}\sim 70-90\units{\msun}$. Thus, even among models tuned to match 47~Tuc’s present-day global properties, the most likely outcome is that the cluster does not retain an IMBH in the classical sense; instead, it retains a comparatively massive stellar-origin BH and ejects the most massive hierarchical-merger products.

Taken together, these results show that 47\,Tuc-like clusters
efficiently produce black holes in the upper mass gap through
hierarchical mergers, but preferentially eject them into the field
via gravitational-wave recoil rather than retaining them. The cluster
itself retains only a $\sim 45 - 70\,\mathrm{M}_\odot$ BH after
$\sim 1 - 3$ merger generations, while contributing heavier
mass-gap remnants ($\sim 90 - 140\,\mathrm{M}_\odot$) to the
gravitational-wave source population without sustained growth toward higher
IMBH scales.

\subsection{IMBH mass in seeded models}\label{sec:seeded_results}

We now examine the extended-IMF (\texttt{S}) models, in which a population
of primordial BH seeds above the (P)PISN gap is injected at $t = 0$
(Sec.~\ref{sec:seed_models}); the results are shown in Table~\ref{tab:tableModel_S}. The analogue selection criteria of
Sec.~\ref{sec:clusterMassRadius} are applied identically to the seeded
grid; the qualifying seeded analogues are \texttt{M04KZ1S}, \texttt{M05KZ1S}, \texttt{M04KZ2S}, \texttt{M05KZ2S}, \texttt{M05HZ1S} and \texttt{M05HZ2S}, which share the same initial cluster properties
($M_{\rm cl,i} = 2 \times 10^6\,\mathrm{M}_\odot$,
$\rho_{\rm h,i} = 3 \times 10^4$ and $10^5\,\mathrm{M}_\odot\,
\mathrm{pc}^{-3}$) as the baseline analogues.

In contrast to the baseline models, the seeded analogues show a strongly bimodal distribution of retained BH masses (see Fig.\,\ref{fig:mIMBHspectra}); when the retained spin is also considered, the distribution becomes trimodal in the joint IMBH (mass, spin) plane, as discussed in Section\ref{subsec:imbh_spin} and shown in Fig.\ref{fig:mIMBH-chiIMBH}. In most
realisations the injected seeds are ejected -- either dynamically or
via gravitational-wave recoil -- and the most massive retained BH reverts to the
stellar-mass channel ($M_{\rm IMBH} \sim 50 - 60\,\mathrm{M}_\odot$,
comparable to the baseline). In a significant minority, however, a
massive seed survives to the present epoch, producing $M_{\rm IMBH}
\sim 300 - 600\,\mathrm{M}_\odot$. This bimodality in the retained mass is reflected in
the large gap between the median and 90th-percentile values: for the
lower-metallicity analogues (\texttt{M04KZ1S}, \texttt{M05KZ1S}, \texttt{M05HZ1S}) the medians are $\sim
60 - 370\,\mathrm{M}_\odot$ with 90th percentiles of $\sim
500 - 1100\,\mathrm{M}_\odot$, while for the higher-metallicity
analogues (\texttt{M04KZ2S}, \texttt{M05KZ2S}, \texttt{M05HZ2S}) the medians are only $\sim
46\,\mathrm{M}_\odot$ but the 90th percentiles reach $\sim
500 - 800\,\mathrm{M}_\odot$ (Table~\ref{tab:tableModel_S}).

The physical origin of this bimodality is the sharp mass threshold
for recoil safety. Seeds with $M_{\rm BH} \gtrsim
450\,\mathrm{M}_\odot$ merge with typical stellar-mass BH companions
($\sim 20 - 30\,\mathrm{M}_\odot$) at extreme mass ratios ($q
\lesssim 0.05$), producing recoil kicks of only $\sim
5 - 20\,\mathrm{km\,s}^{-1}$, well below the escape speed of 47\,Tuc (at the time of merger; current being $v_{\rm esc} \sim 50 - 60\,\mathrm{km\,s}^{-1}$). Such seeds
are therefore retained through multiple mergers and can grow further.
Lighter seeds ($M_{\rm BH} \sim 130 - 300\,\mathrm{M}_\odot$)
occupy a more precarious regime: their first merger with a
stellar-mass companion produces a mass ratio $q \sim 0.07 - 0.23$
and a remnant spin $\chi_{\rm rem} \sim 0.3 - 0.7$, yielding
recoil kicks that range from $\sim 10$ to $\sim
100\,\mathrm{km\,s}^{-1}$ depending on spin orientation. Retention
is therefore stochastic, and those that survive acquire
non-negligible spin that increases the recoil risk for subsequent
mergers. Dynamical three-body ejection provides a parallel removal
channel that progressively scours the lighter seeds over $\sim
10^2\,\mathrm{Myr}$ \citep{Martinez2026}.

The most massive ejected BHs in the seeded analogues are dramatically
heavier than in the baseline case, with median $M_{\rm IMBH,ej} \sim
1100 - 1600\,\mathrm{M}_\odot$. These ejected masses scale with metallicity: 
the $Z = 0.003$ analogues produce $M_{\rm IMBH,ej} \sim 1500\units{\msun}$, while the $Z = 0.007$ analogues
produce $M_{\rm IMBH,ej} \sim 1100\,\mathrm{M}_\odot$,
consistent with the lower seed masses at higher metallicity.

The typical end state for a seeded 47\,Tuc-analogue cluster is
therefore one in which most IMF-origin seeds have been removed and
the surviving configuration is either (i) a single massive retained
seed with $M_{\rm BH} \gtrsim 450\,\mathrm{M}_\odot$ that has grown
modestly through subsequent stellar-mass mergers, (ii) a lower-mass
remnant of $\sim 130 - 300\,\mathrm{M}_\odot$ that avoided ejection
through favourable merger geometries, or (iii) a binary IMBH (e.g.\
$150 + 150\,\mathrm{M}_\odot$ or $300 + 150\,\mathrm{M}_\odot$) if
the last two surviving seeds have not yet merged or been disrupted by
the present epoch,  consistent with the rotating core found in \cite{Bellini2017ApJ}. For the baseline analogues, all retained masses
lie well below the $578\,\mathrm{M}_\odot$ upper limit of \cite{DellaCroce2024}. For the seeded analogues, the median retained
masses are also consistent with this limit, but $\sim 10\%$ of
realisations 
produce retained masses $\gtrsim 500\,\mathrm{M}_\odot$ that
approach or exceed it, underscoring the sensitivity of the
outcome to whether one massive seed survives the dynamical scouring
process.

This highlights a key distinction: while hierarchical mergers alone fail to produce IMBHs in 47 Tuc-like environments, the presence of sufficiently massive primordial seeds ($\gtrsim 450\,\mathrm{M}_\odot$) can bypass the recoil barrier. In their absence, the cluster outcome is instead dominated by mass-gap black holes and the ejection of the most massive merger remnants.

\subsection{IMBH spin}
\label{subsec:imbh_spin}

For each realisation, we also record the dimensionless spin of the most massive BH retained at the present time, $\chi_{\rm IMBH}$ (Table~\ref{tab:tableModel} and~\ref{tab:tableModel_S}).  In our models, the largest retained BH is typically a first- to few-generation merger product, so its spin primarily reflects hierarchical assembly rather than an assumed natal-spin distribution.  Across the full grid, we find $\chi_{\rm IMBH}\sim 0.5 - 0.75$, with most models clustering around $\chi_{\rm IMBH}\simeq 0.6 - 0.7$. Furthermore, for all unseeded models,  we observe no $\chi_\mathrm{IMBH} - M_\mathrm{IMBH}$ dependence (Fig.~\ref{fig:mIMBH-chiIMBH}). These are consistent with the generic expectation that BH--BH merger remnants acquire moderate-to-high spins, with $\chi_{\rm rem}\sim 0.7$ for comparable-mass mergers and somewhat lower values for more unequal mass ratios, even if progenitor BHs are born with low spins. Although the scatter is substantial, $\chi_{\rm IMBH}$ shows a mild trend with the typical merger generation and, indirectly, with the initial escape speed.  Models with lower $\nthG$ (i.e.\ where the most massive retained BH is usually a first-generation remnant) tend to populate the lower end of the spin range, while models that more frequently retain second- or third-generation remnants approach $\chi_{\rm IMBH}\simeq 0.7 - 0.75$ (Table~\ref{tab:tableModel}).  Physically, this reflects the fact that additional merger steps both drive remnants toward the characteristic merger-remnant spin scale and preferentially select survivors from merger sequences with lower recoil and more favourable dynamical histories. The breadth of this range arises because, in our baseline models
($\chi_{\rm natal} = 0$ for all first-generation BHs), the most
massive retained BH is, in some realisations (out of 1000 re-runs for each unique model), still a first-generation
remnant ($\chi = 0$), while in others it is a merger product with
$\chi_{\rm rem} \simeq 0.7$; the population-level scatter therefore
reflects the stochastic merger history rather than an intrinsic spin
distribution. For the subset of 47~Tuc-analogue models (Sec.~\ref{sec:clusterMassRadius}), the retained spins are tightly clustered around $\chi_{\rm IMBH}\simeq 0.62 - 0.67$. Thus, if 47~Tuc retains a single dominant BH at the present time, we expect it to have a moderate spin, $\chi_{\rm IMBH}\sim 0.65$, characteristic of a limited (one- to a few-step) hierarchical merger history.

For the seeded analogues (Sec.~\ref{sec:seeded_results}), the spin
distributions are systematically broader and shifted to lower mean
values compared to the baseline (Table~\ref{tab:tableModel_S}).
Across all six seeded analogues, the mean spin ranges from
$\chi_{\rm IMBH} = 0.46 \pm 0.32$ (\texttt{M05HZ1S}) to
$0.67 \pm 0.19$ (\texttt{M04KZ2S}, \texttt{M05KZ2S}), compared
with $\chi_{\rm IMBH} \simeq 0.62 - 0.67$ for the corresponding
baseline analogues (Table~\ref{tab:tableModel}). The
large standard deviations reflect the same bimodality in the retained
mass distributions: in realisations where all seeds are ejected, the
most massive retained BH is a hierarchical-merger product with
$\chi \sim 0.65$, indistinguishable from the baseline. In realisations where a massive seed ($M_{\rm BH} \gtrsim 450\,\mathrm{M}_\odot$, 
$\chi_{\rm natal} = 0$) survives, its spin remains low even after several mergers 
with stellar-mass companions, because the extreme mass ratio ($q \lesssim 0.05$) 
transfers very little angular momentum to the remnant and produces recoil kicks well below the cluster escape speed 
even at later merger generations. For instance, a
$500\,\mathrm{M}_\odot$ seed merging with a $25\,\mathrm{M}_\odot$
BH acquires $\chi_{\rm rem} \lesssim 0.1$; even after $\sim 5 - 10$
such mergers the spin remains $\chi \lesssim 0.2 - 0.3$.
Alternatively, the massive seed can merge with another seed in a
comparable-mass merger, leading to a remnant that is both high mass
and high spin. This trimodality in the $(M_{\rm IMBH},
\chi_{\rm IMBH})$ space can be seen in Fig.~\ref{fig:mIMBH-chiIMBH},
where seeded models cluster in three distinct regions: a high-spin,
low-mass regime (where all seeds have been ejected and the most
massive BH is a merger product of stellar-mass BHs), a low-spin,
high-mass regime (where the most massive BH is a seed that has
grown through mergers with stellar-mass BHs), and a high-spin,
high-mass regime (where the most massive BH is the product of a
comparable-mass seed--seed merger). This spin--mass correlation
constitutes a potential diagnostic: a central BH in 47~Tuc with
$M_{\rm BH} \gtrsim 300\,\mathrm{M}_\odot$ and low spin ($\chi
\lesssim 0.3$) would favour the primordial-seed origin, whereas
$M_{\rm BH} \sim 50 - 100\,\mathrm{M}_\odot$ with $\chi \sim
0.65$ would point to hierarchical mergers from the standard
stellar-mass channel. This spin--mass distinction provides an observational discriminator 
for gravitational-wave sources originating from dense clusters, 
testable with current LIGO-Virgo-KAGRA observations for merging 
systems and, in the future, with next-generation detectors such as 
the Einstein Telescope \citep{ET2010}, Cosmic Explorer 
\citep{CE2017}, and LISA \citep{LISA2017, LISA2023} for both merging and 
in-band inspiralling systems in dense clusters.

\section{Discussion}
\label{sec:discussion}

Our results show that hierarchical BH--BH mergers in 47~Tuc-like clusters efficiently populate the upper pair-instability mass gap but fail to produce classical IMBHs ($\gtrsim 10^{2\text{--}3}\units{\msun}$) through mergers alone. The key physical reason is that the cluster escape velocity regulates whether early merger remnants can be retained as seeds for subsequent hierarchical growth \citep{Antonini:2018auk,Fragione:2020nib,Mapelli:2021MNRAS,Chattopadhyay2023MNRAS}. After each BH--BH merger, the remnant receives a gravitational-wave recoil kick whose magnitude depends primarily on the progenitor mass ratio and spin configuration; nearly equal-mass, highly spinning progenitors produce the largest kicks \citep{Lousto:2009,Lousto:2010,Varma2019,Islam2023,Chattopadhyay2023MNRAS}. Even if BHs are born with negligible spin, first-generation merger remnants typically acquire $\chi_{\rm rem} \simeq 0.7 - 0.8$, increasing the susceptibility of second- and third-generation remnants to ejection. If a remnant survives the first few generations and grows to dominate the BH population, subsequent mergers occur at progressively smaller mass ratios $q=m_2/m_1\le1$, which reduces both the recoil kick and the remnant spin, thereby improving retention \citep[see Section~3.1.2 of][]{Chattopadhyay2023MNRAS}. This self-regulation makes retention of the earliest merger products critical for reaching bona-fide IMBH masses.

Quantitatively, Fig.~10 of \citet{Chattopadhyay2023MNRAS} shows that clusters with $v_{\rm esc} \lesssim 200\,{\rm km\,s^{-1}}$ typically eject the growing remnant within the first few mergers, whereas $v_{\rm esc} \gtrsim 400\,{\rm km\,s^{-1}}$ makes retention through multiple generations -- and growth to $\gtrsim 10^3\units{\msun}$ -- much more likely. The intermediate regime, $200\,{\rm km\,s^{-1}} \lesssim v_{\rm esc} \lesssim 400\,{\rm km\,s^{-1}}$, is transitional and can yield orders-of-magnitude variation in the retained remnant mass depending on stochastic merger sequences and recoil realisations. In our grid, the initial escape velocities are $v_{\rm esc,h,i} \simeq 63 - 171\,{\rm km\,s^{-1}}$ (Table~\ref{tab:tableModel} and~\ref{tab:tableModel_S}), placing the models squarely in the low-$v_{\rm esc}$ regime.

Adding primordial BH seeds above the (P)PISN gap modifies this picture qualitatively. The key distinction is the sharp mass threshold for recoil safety: seeds above $\sim 450\units{\msun}$ merge at extreme mass ratios that produce negligible kicks \citep{LoustoKick2011}, while lighter seeds are stochastically removed through a combination of gravitational-wave recoil and dynamical scouring on a timescale of $\sim 10^2\,$Myr. The resulting bimodality -- in which the cluster either retains a massive primordial seed or reverts to the stellar-mass channel -- means that the outcome for any individual cluster is highly sensitive to the stochastic high-mass end of the seed population; furthermore, the joint mass--spin distribution is trimodal, with low-spin survivors arising from seed--stellar-mass mergers and high-spin, high-mass remnants from comparable-mass seed--seed mergers. The robust survival criterion of $M_{\rm BH} \gtrsim 10\times$ the upper edge of the (P)PISN gap, i.e.\ $\gtrsim 450\units{\msun}$ \citep[Section~3.2.4 of][]{Chattopadhyay2023MNRAS}, is consistent with our seeded-analogue results.

% Adding primordial BH seeds above the (P)PISN gap modifies this picture qualitatively. The key distinction is the sharp mass threshold for recoil safety: seeds above $\sim 450\units{\msun}$ merge at extreme mass ratios that produce negligible kicks \citep{LoustoKick2011}, while lighter seeds are stochastically removed through a combination of gravitational-wave recoil and dynamical scouring on a timescale of $\sim 10^2\,$Myr. The resulting bimodality -- in which the cluster either retains a massive primordial seed or reverts to the stellar-mass channel -- means that the outcome for any individual cluster is highly sensitive to the stochastic high-mass end of the seed population. The robust survival criterion of $M_{\rm BH} \gtrsim 10\times$ the upper edge of the (P)PISN gap, i.e.\ $\gtrsim 450\units{\msun}$ \citep[Section~3.2.4 of][]{Chattopadhyay2023MNRAS}, is consistent with our seeded-analogue results.

Our results also provide a simple interpretation of the emerging gravitational-wave black hole population: globular clusters in the low escape-speed regime can efficiently repopulate the pair-instability mass gap through hierarchical mergers, while failing to produce intermediate-mass black holes due to gravitational-wave recoil. This implies that the presence of mass-gap black holes in gravitational-wave data does not, by itself, require an efficient pathway to moderately massive ($>10^3\units{\msun}$) IMBH formation. Current LVK analyses find that the merger rate of systems with at least one component above $\sim 45\units{\msun}$ is $\sim 0.1 - 1\,{\rm Gpc^{-3}\,yr^{-1}}$ \citep{LIGOScientific:2025slb}, and our models predict that massive globular clusters are net exporters of such objects. A quantitative rate estimate from the ejected population, and its 
detectability with current and future gravitational-wave detectors is deferred to future work.

The spin--mass correlation described in Section~\ref{subsec:imbh_spin} constitutes a potential observational diagnostic: a central BH in 47~Tuc with $M_{\rm BH} \gtrsim 300\units{\msun}$ would favour the primordial-seed origin, with a high spin ($\chi \sim 0.68$) if it had merged with other seeds or, otherwise, a low spin ($\chi \lesssim 0.3$). However, a $M_{\rm BH} \sim 50 - 100\units{\msun}$ with $\chi \sim 0.65$ would point to hierarchical mergers from the standard stellar-mass channel. Future gravitational-wave observations with next-generation 
ground-based detectors such as the Einstein Telescope and Cosmic Explorer will detect 
mergers out to cosmological distances with sufficient spin 
precision to test this prediction, while LISA 
will constrain the intermediate-mass range directly through 
millihertz-band signals from clusters in the Local Group.

These results are consistent with the stringent dynamical upper limits on any central BH in 47~Tuc \citep{DellaCroce2024}, and support an interpretation in which 47~Tuc's central potential is dominated by a dark-remnant subsystem rather than a single, long-lived IMBH \citep[e.g.][]{HenaultBrunet2020}.

\subsection{Robustness to additional physical processes}
\label{sec:caveats}

Our retained BH masses are conservative lower limits because we neglect mass growth through stellar consumption and do not model stellar-merger seed formation. We can estimate the stellar-consumption rate via gravitational focusing. For a BH of mass $M_\bullet$ in a core of number density $n$ and velocity dispersion $v$, the encounter rate is
\begin{equation}
\Gamma \simeq n\,\sigma\,v,\qquad
\sigma \simeq \frac{2\pi G M_\bullet r_t}{v^2},\qquad
r_t \simeq R_\ast\left(\frac{M_\bullet}{m_\ast}\right)^{1/3},
\end{equation}
where $r_t$ is the tidal radius for a star of mass $m_\ast$ and radius $R_\ast$. For representative old-cluster values ($M_\bullet \simeq 50\units{\msun}$, $m_\ast \simeq 0.8\units{\msun}$, $R_\ast \simeq 1\,R_\odot$, $v \simeq 10 - 12~\mathrm{km~s^{-1}}$), this yields $\sim 10 - 15$ events over 12\,Gyr for $n \sim 10^5$~pc$^{-3}$ and $\sim 100 - 150$ events for $n \sim 10^6$~pc$^{-3}$. With a conservative accretion fraction $f_{\rm acc} \sim 0.1$, the mass growth is $\Delta M_\bullet \sim \mathcal{O}(1)\units{\msun}$ up to $\sim 8 - 12\units{\msun}$; even an optimistic $f_{\rm acc} \sim 1$ gives $\Delta M_\bullet \sim 80 - 120\units{\msun}$. This shifts a $\sim 50 - 70\units{\msun}$ BH to at most $\sim 60 - 190\units{\msun}$, still far below the observational upper limit. We note that this estimate assumes a full loss cone; partial loss-cone refilling may reduce the actual rate by an order of magnitude or more \citep[e.g.][]{Bahcall1976,Lightman1977ApJ,Cohn1978ApJ,Rastello2026}.

The mapping from observed half-light to half-mass radii is not unique and can be biased by mass segregation and stellar-population effects at the $\sim 10 - 20\%$ level \citep{Jordan2004,Mackey2005,Schulman2012}. Similarly, our simplified tidal treatment -- a static potential and circular orbit at $R_{\rm G} = 7.4\,$kpc -- may over- or underestimate late-time mass loss. Time-dependent tides, orbital eccentricity, and disc/bulge shocks can alter the structural evolution \citep{Miholics2014}. However, modest shifts in analogue membership do not move the models out of the low-$v_{\rm esc}$ regime where recoil-limited growth operates. Given that 47~Tuc is on a nearly circular orbit, this effect should be minimal.

\section{Conclusions}
\label{sec:conclusions}

Using the semi-analytical code \texttt{cBHBd} with NR-calibrated remnant prescriptions, we have modelled 80\,000 realisations of 47~Tuc-like clusters
(40\,000 baseline and 40\,000 seeded) to quantify the mass and spin of the most massive BH that can be produced and retained. Our main conclusions are:

\begin{enumerate}

\item Hierarchical BH mergers in 47~Tuc-like clusters ($v_{\rm esc} \lesssim 170\,{\rm km\,s^{-1}}$) are limited to $\sim 1 - 3$ mergers by gravitational-wave recoil. The most massive retained BH has $M_{\rm IMBH} \sim 45 - 70\units{\msun}$ with spin $\chi_{\rm IMBH} \sim 0.65$ -- well below classical IMBH masses and consistent with the $3\sigma$ dynamical upper limit of $578\units{\msun}$ \citep{DellaCroce2024}.

\item These clusters efficiently populate the upper pair-instability mass gap but preferentially eject the most massive merger products ($\sim 70 - 160\units{\msun}$) into the field via gravitational-wave recoil, contributing to the gravitational-wave source population without sustained IMBH growth.

\item When primordial BH seeds above the (P)PISN gap are included, the retained-mass distribution becomes bimodal: in ${\sim}\,90\%$ of realisations all seeds are ejected, but in ${\sim}\,10\%$ a massive seed ($M_{\rm BH} \gtrsim 450\,\mathrm{M}_{\odot}$) survives. The joint mass--spin distribution is trimodal (see Fig.\ref{fig:mIMBH-chiIMBH}): low-spin survivors ($\chi \lesssim 0.3$) arise from seed--stellar-mass mergers, while seed--seed mergers produce high-mass, high-spin remnants ($\chi \sim 0.65 - 0.7$), yielding 90th-percentile retained masses of $\sim 500 - 1100\,\mathrm{M}_{\odot}$.

\item A spin--mass diagnostic distinguishes the two formation channels: $M_{\rm BH} \gtrsim 300\units{\msun}$ with low spin favours a primordial-seed origin, whereas $M_{\rm BH} \sim 50 - 100\units{\msun}$ with $\chi \sim 0.65$ favours hierarchical assembly from the standard stellar-mass channel.

\item 47~Tuc's central potential is most naturally explained by a distributed dark-remnant subsystem -- possibly anchored by one or two surviving primordial seeds -- rather than a single massive ($> 10^3\units{\msun}$) IMBH.

\end{enumerate}

Since 47~Tuc is among the most massive Milky Way globular clusters, the difficulty of growing an IMBH through hierarchical mergers here implies that this channel is even less effective in typical, lower-mass clusters.

\section*{Acknowledgements}
DC and DMP acknowledge the Kavli Institute for Theoretical Physics (KITP) for hospitality and support during the completion of this work during the program ``Stellar-Mass Black Holes at the Nexus of Optical, X-ray, and Gravitational Wave Surveys'' (2025), supported in part by the National Science Foundation under Grant NSF PHY-2309135. DC thanks the Gordon and Betty Moore Foundation for funding this research through Grant GBMF12341 and previously STFC grant ST/V005618/1. DMP acknowledges support from the Deutsche Forschungsgemeinschaft (DFG, German Research Foundation) through project number 546850815 (acronym: DoBlack) and under Germany's Excellence Strategy EXC 2181/1 - 390900948 (the Heidelberg STRUCTURES Excellence Cluster). FA is supported by the UK’s Science and Technology Facilities Council grant
ST/V005618/1. MG acknowledges   the grants PID2024-155720NB-I00, CEX2024-001451-M funded by MCIN/AEI/10.13039/501100011033.
FFP acknowledges the  “la Caixa” Foundation (ID100010434) for financial support in the form of a Doctoral INPhINIT fellowship (fellowship code LCF/BQ/DI23/11990067).

% \dani{once published, we have to write here https://www.kitp.ucsb.edu/report-publication} \debatri{check}
\section*{Data Availability}

The data utilized for this work will be freely available  upon reasonable request to the corresponding author(s). The code \cbhbd will be made available in the public domain GitHub \url{https://github.com/cBHBd/cBHBd}. %\debatri{is it now readily public ?}\dani{will be published soon, but surely before submitting this letter}

\newlength{\tabcolcust}
\setlength{\tabcolcust}{9.5pt}

\begin{table*}
\caption{Models without IMBH seeds. Summary of results from our simulations of 47~Tuc-like clusters. The columns are, from left to right: the name of the model, the initial cluster mass ($M_\mathrm{cl, i}$), the initial density within the half-mass radius ($\rho_\mathrm{h,i}$), the metallicity ($Z$), the chosen IMF, the initial escape velocity ($v_\mathrm{esc,h,i}$), the final mass ($M_\mathrm{cl,f}$), the final half-mass radius ($r_\mathrm{h,f}$), the mass of the most massive BH retained in the cluster ($M_\mathrm{IMBH}$; shown as median, 10th, and 90th percentile), the mass of the most massive BH ejected from the cluster ($M_\mathrm{IMBH, ej}$; shown as median, 10th, and 90th percentile), the generation of the most massive retained BH ($\nthG$, shown as mean and standard deviation), and the spin of the most massive retained BH ($\chi_\mathrm{IMBH}$, shown as mean and standard deviation). The 47~Tuc analogue models are marked in bold letters.}
\label{tab:tableModel}
\centering
\begin{tabular}{l@{\hspace{\tabcolcust}}c@{\hspace{\tabcolcust}}c@{\hspace{\tabcolcust}}c@{\hspace{\tabcolcust}}c@{\hspace{\tabcolcust}}c@{\hspace{\tabcolcust}}c@{\hspace{\tabcolcust}}c@{\hspace{\tabcolcust}}c@{\hspace{\tabcolcust}}c@{\hspace{\tabcolcust}}c@{\hspace{\tabcolcust}}c}
\hline\hline
Model & $M_\mathrm{cl,i}$  & $\rho_\mathrm{h,i}$ & $Z$ & IMF & $v_\mathrm{esc,h,i}$ & $M_\mathrm{cl,f}$ & $r_\mathrm{h,f}$ & $M_\mathrm{IMBH}$ & $M_\mathrm{IMBH, ej}$   & $\nthG$ & $\chi_\mathrm{IMBH}$ \\
name & (M$_\odot$) & (M$_\odot$\,pc$^{-3}$) & & & (km/s) & (M$_\odot$) & (pc) &(M$_\odot$)& (M$_\odot$) && \\
\hline
\hline
M01KZ1 & $2\times 10^6$ & $1\times 10^3$ & 0.003 & K & 63 &$1.12\times 10^6$ & $16.2$ & $59^{+8}_{-22}$ & $96^{+6}_{-16}$ & $1.8 \pm 0.4$ & $0.52 \pm 0.30$ \\
M02KZ1 & $2\times 10^6$ & $3\times 10^3$ & 0.003 & K & 76 &$1.10\times 10^6$ & $12.9$ & $61^{+5}_{-24}$ & $99^{+14}_{-6}$ & $1.9 \pm 0.4$ & $0.59 \pm 0.24$ \\
M03KZ1 & $2\times 10^6$ & $1\times 10^4$ & 0.003 & K & 92 &$1.09\times 10^6$ & $10.1$ & $61^{+5}_{-17}$ & $102^{+21}_{-5}$ & $1.9 \pm 0.3$ & $0.64 \pm 0.18$ \\
\textbf{M04KZ1} & $2\times 10^6$ & $3\times 10^4$ & 0.003 & K & 111 &$1.07\times 10^6$ & $8.1$ & $62^{+8}_{-19}$ & $114^{+14}_{-13}$ & $2.0 \pm 0.3$ & $0.64 \pm 0.17$ \\
\textbf{M05KZ1} & $2\times 10^6$ & $1\times 10^5$ & 0.003 & K & 136 &$1.04\times 10^6$ & $6.3$ & $70^{+0}_{-22}$ & $124^{+10}_{-17}$ & $2.0 \pm 0.3$ & $0.66 \pm 0.15$ \\
M06KZ1 & $4\times 10^6$ & $1\times 10^3$ & 0.003 & K & 79 &$2.28\times 10^6$ & $18.0$ & $64^{+2}_{-6}$ & $100^{+18}_{-5}$ & $2.0 \pm 0.2$ & $0.67 \pm 0.10$ \\
M07KZ1 & $4\times 10^6$ & $3\times 10^3$ & 0.003 & K & 95 &$2.26\times 10^6$ & $14.1$ & $64^{+3}_{-4}$ & $103^{+22}_{-4}$ & $2.0 \pm 0.2$ & $0.68 \pm 0.06$ \\
M08KZ1 & $4\times 10^6$ & $1\times 10^4$ & 0.003 & K & 116 &$2.23\times 10^6$ & $10.9$ & $64^{+6}_{-3}$ & $121^{+9}_{-18}$ & $2.1 \pm 0.3$ & $0.70 \pm 0.04$ \\
M09KZ1 & $4\times 10^6$ & $3\times 10^4$ & 0.003 & K & 140 &$2.19\times 10^6$ & $8.7$ & $64^{+26}_{-3}$ & $126^{+8}_{-8}$ & $2.2 \pm 0.4$ & $0.71 \pm 0.06$ \\
M10KZ1 & $4\times 10^6$ & $1\times 10^5$ & 0.003 & K & 171 &$2.15\times 10^6$ & $6.8$ & $63^{+31}_{-1}$ & $131^{+22}_{-7}$ & $2.3 \pm 0.5$ & $0.72 \pm 0.07$ \\
\hline
M01HZ1 & $2\times 10^6$ & $1\times 10^3$ & 0.003 & H & 63 &$9.80\times 10^5$ & $19.4$ & $43^{+22}_{-6}$ & $90^{+11}_{-23}$ & $1.5 \pm 0.5$ & $0.36 \pm 0.34$ \\
M02HZ1 & $2\times 10^6$ & $3\times 10^3$ & 0.003 & H & 76 &$9.71\times 10^5$ & $15.6$ & $56^{+10}_{-19}$ & $97^{+5}_{-12}$ & $1.7 \pm 0.5$ & $0.48 \pm 0.31$ \\
M03HZ1 & $2\times 10^6$ & $1\times 10^4$ & 0.003 & H & 92 &$9.57\times 10^5$ & $12.3$ & $58^{+7}_{-22}$ & $100^{+20}_{-5}$ & $1.8 \pm 0.4$ & $0.55 \pm 0.28$ \\
M04HZ1 & $2\times 10^6$ & $3\times 10^4$ & 0.003 & H & 111 &$9.40\times 10^5$ & $10.0$ & $59^{+8}_{-22}$ & $103^{+22}_{-5}$ & $1.9 \pm 0.4$ & $0.60 \pm 0.23$ \\
\textbf{M05HZ1} & $2\times 10^6$ & $1\times 10^5$ & 0.003 & H & 136 &$9.16\times 10^5$ & $8.0$ & $61^{+9}_{-23}$ & $120^{+12}_{-17}$ & $1.9 \pm 0.4$ & $0.62 \pm 0.20$ \\
M06HZ1 & $4\times 10^6$ & $1\times 10^3$ & 0.003 & H & 79 &$2.01\times 10^6$ & $21.3$ & $63^{+3}_{-26}$ & $98^{+5}_{-13}$ & $1.9 \pm 0.3$ & $0.60 \pm 0.22$ \\
M07HZ1 & $4\times 10^6$ & $3\times 10^3$ & 0.003 & H & 95 &$2.00\times 10^6$ & $16.9$ & $64^{+3}_{-8}$ & $101^{+21}_{-4}$ & $2.0 \pm 0.2$ & $0.66 \pm 0.13$ \\
M08HZ1 & $4\times 10^6$ & $1\times 10^4$ & 0.003 & H & 116 &$1.97\times 10^6$ & $13.2$ & $64^{+3}_{-5}$ & $115^{+13}_{-14}$ & $2.0 \pm 0.2$ & $0.68 \pm 0.05$ \\
M09HZ1 & $4\times 10^6$ & $3\times 10^4$ & 0.003 & H & 140 &$1.94\times 10^6$ & $10.7$ & $64^{+6}_{-5}$ & $123^{+8}_{-15}$ & $2.1 \pm 0.3$ & $0.70 \pm 0.05$ \\
M10HZ1 & $4\times 10^6$ & $1\times 10^5$ & 0.003 & H & 171 &$1.91\times 10^6$ & $8.5$ & $63^{+26}_{-4}$ & $128^{+19}_{-7}$ & $2.2 \pm 0.4$ & $0.71 \pm 0.06$ \\
\hline
M01KZ2 & $2\times 10^6$ & $1\times 10^3$ & 0.007 & K & 63 &$1.10\times 10^6$ & $15.2$ & $45^{+3}_{-18}$ & $70^{+3}_{-8}$ & $1.8 \pm 0.4$ & $0.57 \pm 0.26$ \\
M02KZ2 & $2\times 10^6$ & $3\times 10^3$ & 0.007 & K & 76 &$1.09\times 10^6$ & $11.9$ & $45^{+2}_{-13}$ & $72^{+13}_{-3}$ & $1.9 \pm 0.3$ & $0.63 \pm 0.18$ \\
M03KZ2 & $2\times 10^6$ & $1\times 10^4$ & 0.007 & K & 92 &$1.08\times 10^6$ & $9.2$ & $45^{+5}_{-9}$ & $74^{+16}_{-3}$ & $2.0 \pm 0.2$ & $0.66 \pm 0.13$ \\
\textbf{M04KZ2} & $2\times 10^6$ & $3\times 10^4$ & 0.007 & K & 111 &$1.06\times 10^6$ & $7.3$ & $46^{+4}_{-9}$ & $87^{+8}_{-13}$ & $2.0 \pm 0.3$ & $0.67 \pm 0.11$ \\
\textbf{M05KZ2} & $2\times 10^6$ & $1\times 10^5$ & 0.007 & K & 136 &$1.03\times 10^6$ & $5.7$ & $45^{+5}_{-9}$ & $93^{+4}_{-9}$ & $2.1 \pm 0.3$ & $0.67 \pm 0.12$ \\
M06KZ2 & $4\times 10^6$ & $1\times 10^3$ & 0.007 & K & 79 &$2.25\times 10^6$ & $17.1$ & $47^{+2}_{-4}$ & $72^{+14}_{-4}$ & $2.0 \pm 0.1$ & $0.68 \pm 0.07$ \\
M07KZ2 & $4\times 10^6$ & $3\times 10^3$ & 0.007 & K & 95 &$2.23\times 10^6$ & $13.2$ & $47^{+2}_{-2}$ & $80^{+11}_{-9}$ & $2.0 \pm 0.2$ & $0.69 \pm 0.04$ \\
M08KZ2 & $4\times 10^6$ & $1\times 10^4$ & 0.007 & K & 116 &$2.21\times 10^6$ & $10.1$ & $46^{+4}_{-1}$ & $89^{+5}_{-9}$ & $2.1 \pm 0.3$ & $0.69 \pm 0.04$ \\
M09KZ2 & $4\times 10^6$ & $3\times 10^4$ & 0.007 & K & 140 &$2.17\times 10^6$ & $8.0$ & $46^{+22}_{-1}$ & $92^{+5}_{-5}$ & $2.3 \pm 0.4$ & $0.72 \pm 0.07$ \\
M10KZ2 & $4\times 10^6$ & $1\times 10^5$ & 0.007 & K & 171 &$2.13\times 10^6$ & $6.2$ & $50^{+22}_{-6}$ & $96^{+18}_{-5}$ & $2.5 \pm 0.5$ & $0.74 \pm 0.08$ \\
\hline
M01HZ2 & $2\times 10^6$ & $1\times 10^3$ & 0.007 & H & 63 &$9.69\times 10^5$ & $18.5$ & $40^{+8}_{-13}$ & $67^{+5}_{-18}$ & $1.6 \pm 0.5$ & $0.41 \pm 0.34$ \\
M02HZ2 & $2\times 10^6$ & $3\times 10^3$ & 0.007 & H & 76 &$9.62\times 10^5$ & $14.7$ & $43^{+4}_{-17}$ & $71^{+3}_{-7}$ & $1.8 \pm 0.4$ & $0.52 \pm 0.29$ \\
M03HZ2 & $2\times 10^6$ & $1\times 10^4$ & 0.007 & H & 92 &$9.49\times 10^5$ & $11.6$ & $44^{+4}_{-18}$ & $73^{+16}_{-3}$ & $1.9 \pm 0.4$ & $0.59 \pm 0.23$ \\
M04HZ2 & $2\times 10^6$ & $3\times 10^4$ & 0.007 & H & 111 &$9.32\times 10^5$ & $9.4$ & $45^{+6}_{-17}$ & $77^{+14}_{-5}$ & $1.9 \pm 0.3$ & $0.62 \pm 0.20$ \\
\textbf{M05HZ2} & $2\times 10^6$ & $1\times 10^5$ & 0.007 & H & 136 &$9.08\times 10^5$ & $7.5$ & $46^{+4}_{-14}$ & $88^{+7}_{-14}$ & $2.0 \pm 0.3$ & $0.64 \pm 0.17$ \\
M06HZ2 & $4\times 10^6$ & $1\times 10^3$ & 0.007 & H & 79 &$1.98\times 10^6$ & $20.6$ & $46^{+2}_{-20}$ & $70^{+3}_{-12}$ & $1.9 \pm 0.3$ & $0.61 \pm 0.21$ \\
M07HZ2 & $4\times 10^6$ & $3\times 10^3$ & 0.007 & H & 95 &$1.97\times 10^6$ & $16.2$ & $46^{+2}_{-4}$ & $73^{+16}_{-3}$ & $2.0 \pm 0.1$ & $0.68 \pm 0.08$ \\
M08HZ2 & $4\times 10^6$ & $1\times 10^4$ & 0.007 & H & 116 &$1.95\times 10^6$ & $12.6$ & $46^{+2}_{-2}$ & $84^{+8}_{-11}$ & $2.0 \pm 0.2$ & $0.69 \pm 0.02$ \\
M09HZ2 & $4\times 10^6$ & $3\times 10^4$ & 0.007 & H & 140 &$1.93\times 10^6$ & $10.1$ & $46^{+7}_{-2}$ & $90^{+5}_{-9}$ & $2.1 \pm 0.3$ & $0.70 \pm 0.05$ \\
M10HZ2 & $4\times 10^6$ & $1\times 10^5$ & 0.007 & H & 171 &$1.89\times 10^6$ & $8.0$ & $46^{+21}_{-1}$ & $93^{+16}_{-5}$ & $2.2 \pm 0.4$ & $0.71 \pm 0.06$ \\
\hline
\end{tabular}
\end{table*}

\begin{table*}
\caption{Models with IMBH seeds (\texttt{S})}
\label{tab:tableModel_S}
\centering
\begin{tabular}{l@{\hspace{\tabcolcust}}c@{\hspace{\tabcolcust}}c@{\hspace{\tabcolcust}}c@{\hspace{\tabcolcust}}c@{\hspace{\tabcolcust}}c@{\hspace{\tabcolcust}}c@{\hspace{\tabcolcust}}c@{\hspace{\tabcolcust}}c@{\hspace{\tabcolcust}}c@{\hspace{\tabcolcust}}c@{\hspace{\tabcolcust}}c}
% \begin{tabular}{lccccccccccc}
\hline
Model & $M_\mathrm{cl,i}$  & $\rho_\mathrm{h,i}$ & $Z$ & IMF & $v_\mathrm{esc,h,i}$ & $M_\mathrm{cl,f}$ & $r_\mathrm{h,f}$ & $M_\mathrm{IMBH}$ & $M_\mathrm{IMBH, ej}$   & $\nthG$ & $\chi_\mathrm{IMBH}$ \\
name & (M$_\odot$) & (M$_\odot$\,pc$^{-3}$) & & & (km/s) & (M$_\odot$) & (pc) &(M$_\odot$)& (M$_\odot$) && \\
\hline
\hline
M01KZ1S & $2\times 10^6$ & $1\times 10^3$ & 0.003 & K & 63 &$1.15\times 10^6$ & $17.0$ & $455^{+313}_{-125}$ & $1383^{+336}_{-338}$ & $1.3 \pm 0.5$ & $0.22 \pm 0.32$ \\
M02KZ1S & $2\times 10^6$ & $3\times 10^3$ & 0.003 & K & 76 &$1.14\times 10^6$ & $13.5$ & $368^{+254}_{-84}$ & $1493^{+291}_{-391}$ & $1.5 \pm 0.5$ & $0.32 \pm 0.34$ \\
M03KZ1S & $2\times 10^6$ & $1\times 10^4$ & 0.003 & K & 92 &$1.12\times 10^6$ & $10.6$ & $337^{+197}_{-300}$ & $1574^{+241}_{-368}$ & $1.7 \pm 0.6$ & $0.48 \pm 0.32$ \\
\textbf{M04KZ1S} & $2\times 10^6$ & $3\times 10^4$ & 0.003 & K & 111 &$1.10\times 10^6$ & $8.6$ & $61^{+432}_{-24}$ & $1585^{+242}_{-280}$ & $1.9 \pm 1.1$ & $0.55 \pm 0.28$ \\
\textbf{M05KZ1S} & $2\times 10^6$ & $1\times 10^5$ & 0.003 & K & 136 &$1.07\times 10^6$ & $6.7$ & $370^{+738}_{-319}$ & $1415^{+380}_{-621}$ & $2.1 \pm 2.0$ & $0.56 \pm 0.28$ \\
M06KZ1S & $4\times 10^6$ & $1\times 10^3$ & 0.003 & K & 79 &$2.31\times 10^6$ & $18.4$ & $475^{+428}_{-177}$ & $1489^{+280}_{-403}$ & $1.5 \pm 0.5$ & $0.34 \pm 0.34$ \\
M07KZ1S & $4\times 10^6$ & $3\times 10^3$ & 0.003 & K & 95 &$2.29\times 10^6$ & $14.3$ & $361^{+325}_{-324}$ & $1560^{+234}_{-411}$ & $1.8 \pm 0.5$ & $0.51 \pm 0.30$ \\
M08KZ1S & $4\times 10^6$ & $1\times 10^4$ & 0.003 & K & 116 &$2.26\times 10^6$ & $11.1$ & $65^{+775}_{-5}$ & $1593^{+251}_{-304}$ & $2.6 \pm 2.4$ & $0.70 \pm 0.07$ \\
M09KZ1S & $4\times 10^6$ & $3\times 10^4$ & 0.003 & K & 140 &$2.23\times 10^6$ & $8.9$ & $65^{+876}_{-4}$ & $1623^{+280}_{-256}$ & $3.6 \pm 5.5$ & $0.71 \pm 0.07$ \\
M10KZ1S & $4\times 10^6$ & $1\times 10^5$ & 0.003 & K & 171 &$2.18\times 10^6$ & $7.0$ & $70^{+1107}_{-8}$ & $1659^{+417}_{-254}$ & $5.6 \pm 10.0$ & $0.73 \pm 0.09$ \\
\hline
M01HZ1S & $2\times 10^6$ & $1\times 10^3$ & 0.003 & H & 63 &$1.02\times 10^6$ & $20.3$ & $532^{+353}_{-159}$ & $1376^{+331}_{-342}$ & $1.3 \pm 0.5$ & $0.21 \pm 0.32$ \\
M02HZ1S & $2\times 10^6$ & $3\times 10^3$ & 0.003 & H & 76 &$1.01\times 10^6$ & $16.3$ & $430^{+322}_{-122}$ & $1464^{+282}_{-350}$ & $1.4 \pm 0.5$ & $0.25 \pm 0.33$ \\
M03HZ1S & $2\times 10^6$ & $1\times 10^4$ & 0.003 & H & 92 &$9.93\times 10^5$ & $13.0$ & $342^{+295}_{-92}$ & $1560^{+226}_{-345}$ & $1.5 \pm 0.5$ & $0.35 \pm 0.34$ \\
M04HZ1S & $2\times 10^6$ & $3\times 10^4$ & 0.003 & H & 111 &$9.74\times 10^5$ & $10.5$ & $317^{+221}_{-262}$ & $1584^{+214}_{-283}$ & $1.7 \pm 0.5$ & $0.46 \pm 0.33$ \\
\textbf{M05HZ1S} & $2\times 10^6$ & $1\times 10^5$ & 0.003 & H & 136 &$9.49\times 10^5$ & $8.4$ & $63^{+453}_{-26}$ & $1606^{+255}_{-223}$ & $1.7 \pm 0.8$ & $0.46 \pm 0.32$ \\
M06HZ1S & $4\times 10^6$ & $1\times 10^3$ & 0.003 & H & 79 &$2.05\times 10^6$ & $21.8$ & $579^{+471}_{-192}$ & $1442^{+307}_{-394}$ & $1.4 \pm 0.5$ & $0.27 \pm 0.34$ \\
M07HZ1S & $4\times 10^6$ & $3\times 10^3$ & 0.003 & H & 95 &$2.03\times 10^6$ & $17.3$ & $433^{+427}_{-145}$ & $1535^{+253}_{-407}$ & $1.5 \pm 0.5$ & $0.33 \pm 0.34$ \\
M08HZ1S & $4\times 10^6$ & $1\times 10^4$ & 0.003 & H & 116 &$2.01\times 10^6$ & $13.5$ & $354^{+337}_{-308}$ & $1576^{+250}_{-351}$ & $1.7 \pm 0.5$ & $0.48 \pm 0.32$ \\
M09HZ1S & $4\times 10^6$ & $3\times 10^4$ & 0.003 & H & 140 &$1.98\times 10^6$ & $11.0$ & $65^{+641}_{-11}$ & $1608^{+259}_{-248}$ & $2.1 \pm 1.1$ & $0.64 \pm 0.18$ \\
M10HZ1S & $4\times 10^6$ & $1\times 10^5$ & 0.003 & H & 171 &$1.94\times 10^6$ & $8.7$ & $65^{+735}_{-4}$ & $1632^{+339}_{-245}$ & $2.8 \pm 3.3$ & $0.70 \pm 0.08$ \\
\hline
M01KZ2S & $2\times 10^6$ & $1\times 10^3$ & 0.007 & K & 63 &$1.13\times 10^6$ & $15.6$ & $359^{+264}_{-92}$ & $1012^{+222}_{-263}$ & $1.4 \pm 0.5$ & $0.30 \pm 0.34$ \\
M02KZ2S & $2\times 10^6$ & $3\times 10^3$ & 0.007 & K & 76 &$1.11\times 10^6$ & $12.3$ & $330^{+190}_{-153}$ & $1096^{+175}_{-280}$ & $1.7 \pm 0.5$ & $0.47 \pm 0.33$ \\
M03KZ2S & $2\times 10^6$ & $1\times 10^4$ & 0.007 & K & 92 &$1.10\times 10^6$ & $9.6$ & $46^{+502}_{-20}$ & $1126^{+161}_{-240}$ & $2.3 \pm 1.8$ & $0.57 \pm 0.29$ \\
\textbf{M04KZ2S} & $2\times 10^6$ & $3\times 10^4$ & 0.007 & K & 111 &$1.08\times 10^6$ & $7.6$ & $46^{+640}_{-10}$ & $1149^{+202}_{-185}$ & $3.3 \pm 4.4$ & $0.67 \pm 0.19$ \\
\textbf{M05KZ2S} & $2\times 10^6$ & $1\times 10^5$ & 0.007 & K & 136 &$1.05\times 10^6$ & $5.9$ & $46^{+795}_{-6}$ & $1139^{+200}_{-298}$ & $4.2 \pm 6.9$ & $0.67 \pm 0.19$ \\
M06KZ2S & $4\times 10^6$ & $1\times 10^3$ & 0.007 & K & 79 &$2.27\times 10^6$ & $17.3$ & $410^{+336}_{-140}$ & $1061^{+206}_{-290}$ & $1.5 \pm 0.5$ & $0.35 \pm 0.34$ \\
M07KZ2S & $4\times 10^6$ & $3\times 10^3$ & 0.007 & K & 95 &$2.25\times 10^6$ & $13.4$ & $357^{+271}_{-322}$ & $1119^{+179}_{-278}$ & $2.3 \pm 1.3$ & $0.61 \pm 0.26$ \\
M08KZ2S & $4\times 10^6$ & $1\times 10^4$ & 0.007 & K & 116 &$2.23\times 10^6$ & $10.2$ & $48^{+727}_{-3}$ & $1145^{+202}_{-217}$ & $4.1 \pm 5.0$ & $0.72 \pm 0.10$ \\
M09KZ2S & $4\times 10^6$ & $3\times 10^4$ & 0.007 & K & 140 &$2.20\times 10^6$ & $8.1$ & $49^{+1151}_{-4}$ & $1191^{+227}_{-201}$ & $7.4 \pm 11.0$ & $0.74 \pm 0.10$ \\
M10KZ2S & $4\times 10^6$ & $1\times 10^5$ & 0.007 & K & 171 &$2.15\times 10^6$ & $6.3$ & $66^{+1585}_{-21}$ & $1224^{+477}_{-185}$ & $10.3 \pm 16.1$ & $0.76 \pm 0.11$ \\
\hline
M01HZ2S & $2\times 10^6$ & $1\times 10^3$ & 0.007 & H & 63 &$9.94\times 10^5$ & $19.0$ & $402^{+306}_{-111}$ & $980^{+248}_{-228}$ & $1.3 \pm 0.5$ & $0.22 \pm 0.32$ \\
M02HZ2S & $2\times 10^6$ & $3\times 10^3$ & 0.007 & H & 76 &$9.87\times 10^5$ & $15.2$ & $327^{+274}_{-81}$ & $1083^{+183}_{-283}$ & $1.4 \pm 0.5$ & $0.31 \pm 0.34$ \\
M03HZ2S & $2\times 10^6$ & $1\times 10^4$ & 0.007 & H & 92 &$9.73\times 10^5$ & $12.0$ & $308^{+154}_{-150}$ & $1111^{+165}_{-256}$ & $1.7 \pm 0.5$ & $0.46 \pm 0.32$ \\
M04HZ2S & $2\times 10^6$ & $3\times 10^4$ & 0.007 & H & 111 &$9.55\times 10^5$ & $9.7$ & $46^{+379}_{-20}$ & $1132^{+158}_{-195}$ & $1.8 \pm 1.0$ & $0.48 \pm 0.32$ \\
\textbf{M05HZ2S} & $2\times 10^6$ & $1\times 10^5$ & 0.007 & H & 136 &$9.31\times 10^5$ & $7.7$ & $46^{+429}_{-19}$ & $1143^{+179}_{-168}$ & $2.3 \pm 2.3$ & $0.59 \pm 0.25$ \\
M06HZ2S & $4\times 10^6$ & $1\times 10^3$ & 0.007 & H & 79 &$2.01\times 10^6$ & $20.8$ & $454^{+330}_{-137}$ & $967^{+273}_{-239}$ & $1.4 \pm 0.5$ & $0.27 \pm 0.34$ \\
M07HZ2S & $4\times 10^6$ & $3\times 10^3$ & 0.007 & H & 95 &$2.00\times 10^6$ & $16.4$ & $362^{+305}_{-116}$ & $1121^{+166}_{-286}$ & $1.5 \pm 0.5$ & $0.37 \pm 0.34$ \\
M08HZ2S & $4\times 10^6$ & $1\times 10^4$ & 0.007 & H & 116 &$1.98\times 10^6$ & $12.7$ & $330^{+267}_{-303}$ & $1133^{+163}_{-197}$ & $2.1 \pm 1.3$ & $0.58 \pm 0.28$ \\
M09HZ2S & $4\times 10^6$ & $3\times 10^4$ & 0.007 & H & 140 &$1.95\times 10^6$ & $10.2$ & $47^{+573}_{-4}$ & $1143^{+187}_{-169}$ & $2.9 \pm 3.2$ & $0.70 \pm 0.10$ \\
M10HZ2S & $4\times 10^6$ & $1\times 10^5$ & 0.007 & H & 171 &$1.91\times 10^6$ & $8.1$ & $48^{+894}_{-3}$ & $1183^{+212}_{-162}$ & $5.0 \pm 7.8$ & $0.72 \pm 0.08$ \\
\hline
\end{tabular}
\end{table*}

\bibliographystyle{aa}
\bibliography{example} % if your bibtex file is called example.bib

%\begin{thebibliography}{99}
%\end{thebibliography}

%%%%%%%%%%%%%%%%%%%%%%%%%%%%%%%%%%%%%%%%%%%%%%%%%%

%%%%%%%%%%%%%%%%% APPENDICES %%%%%%%%%%%%%%%%%%%%%

\end{document}